\documentclass[prl,floatfix,twocolumn,showpacs]{revtex4}

\usepackage{graphicx}%
\usepackage{dcolumn}
\usepackage{amsmath}
\begin{document}

\preprint{cond-mat/none so far}

\title[Short Title]{Assortative Mixing by Degree Makes a Network
More Unstable}
\author{Markus Brede$^1$}
\author{Sitabhra Sinha$^2$}
\affiliation{%
$^1$CSIRO Centre for Complex Systems, Canberra, ACT 2601, Australia\\
$^2$The Institute of Mathematical Sciences, C.I.T. Campus, Taramani, Chennai - 600 113 India
}%

\date{\today}
\begin{abstract}
We investigate the role of degree correlation among 
nodes on the stability of
complex networks, by studying spectral properties of 
randomly weighted matrices constructed from  directed Erd\"{o}s-R\'enyi and 
scale-free random 
graph models. We focus on the behaviour of the largest real part
of the eigenvalues, $\lambda_\text{max}$, that governs the growth rate
of perturbations about an equilibrium (and hence, determines stability). 
We find that assortative mixing by degree, where nodes with many links connect
preferentially to other nodes with many links, reduces the
stability of networks. In particular, for sparse 
scale-free networks with $N$ nodes, $\lambda_\text{max}$
scales as $N^\alpha$ 
for highly assortative networks, while for disassortative graphs, 
$\lambda_\text{max}$ scales logarithmically with $N$.
This difference may be a possible reason
for the prevalence of disassortative networks in nature.
\end{abstract}
\pacs{89.75.Hc, 05.65.+b, 05.45.-a}
\maketitle

Whether complex systems are more stable than simpler ones has been a 
contentious issue for a long time \cite{McCann,May1}. Stability of
a $N-$dimensional dynamical system
$\dot{\bf x} = -{\bf x}+F({\bf x})$,
(e.g., a network of $N$ nodes, each node $i$ associated with a variable $x_i$
whose time evolution is governed by a general nonlinear
function $F$)
is measured by the rate at which
a small perturbation about an equilibrium state
${\bf x}^*$ 
decays with time. This 
is determined by the largest real part, $\lambda_\text{max}$, of the 
eigenvalues of the Jacobian matrix 
$J_{i j}=-\delta_{i j}+ \partial F_i/\partial x_j |_{\bf{x}^*}$, 
representing the interactions within the system. If $\lambda_\text{max}$ 
is positive,
the effect of a small perturbation will grow exponentially with time, 
and the system will be rapidly dislodged from the equilibrium state.
In the absence of detailed knowledge about the interactions
among the constituent elements of a complex system
which determine $J_{i j}$, 
ensembles of randomly constructed matrices
have been studied \cite{May2}. Results from several studies of 
random matrices, symmetric as well as asymmetric 
\cite{Sommers,Mehta}, show 
that $\lambda_\text{max}\sim \sqrt{N C \sigma^2}$ where $N$ is the 
system size, $C$ is the connectivity (the probability that there is a link
between any pair of nodes in the network) and $\sigma$ is
a measure for interaction strength between the nodes. 
Assuming self regulation, e.g., $J_{ii}=-1$, it follows that a 
transition from stability to instability takes place as the network either
grows in size or becomes more densely (or strongly) connected. This result, 
implying complexity promotes instability, has been shown to be remarkably
robust with respect to various generalisations \cite{Jirsa,Sin1,Sin2}.

There is a direct correspondence between the nature of the matrices $J$
and the structure of the underlying directed network, because two 
nodes $i, j$ are connected
iff $J_{i j}\not= 0$. Clearly, the random matrices investigated in these 
previous studies correspond to Erd\"{o}s-R\'enyi (ER) random graphs 
\cite{Erdos}.
Recently, it has been pointed out in a series of studies that many real 
world networks differ from these purely random graphs 
\cite{BarabRev,NewmanRev}, e.g.,  
in having highly skewed degree distributions that
typically follow a power law with the exponent $\gamma$ 
between $-2$ and $-3$ \cite{Barabasi}. 
Previous studies have investigated 
spectral properties of undirected binary
graphs with power law degree distribution 
\cite{Goh,Farkas,Dorogovtsev}. 
However, there are further detailed features that make real world 
networks appear fundamentally different from random graphs. 
In particular, in almost all biological networks, nodes of high degree 
tend to avoid being connected to other highly connected nodes, i.e. these 
networks show disassortative mixing \cite{Newman02,Newman03}. 
Understanding why networks exhibit these patterns is one of the challenges 
of current research \cite{Agenda}.

In this paper we show that assortativity can significantly
influence network stability. 
For this, we study spectral properties of randomly weighted
matrix ensembles into which we 
introduce correlations between the matrix elements $J_{ij}$ that reflect 
graph properties found in real world networks.
We first examine the growth of $\lambda_{\text{max}}$ with network size in 
matrix ensembles corresponding to graphs with Poisson degree 
distributions. Increasing the assortativity of such graphs 
results in increasing
the largest eigenvalue, implying decrease in stability.
Next, we study the behaviour of growing networks with scale free
degree distribution. We observe the eigenvalue distribution 
(that departs significantly from the spectra of the corresponding 
adjacency matrices)
and determine the scaling of $\lambda_{\text{max}}$ with network size. 
Again we find that assortative mixing leads to a 
loss of stability.
Our results seem to imply that 
the disassortative mixing seen in biological networks could be a 
consequence of evolutionary optimisation of these networks to minimise 
the effect of dynamical fluctuations.

Every directed graph can be represented by its adjacency matrix $A_{ij}$, 
where $A_{i j}=1$ if there is a link from $j$ to $i$; $=0$ otherwise. 
Following May \cite{May2}, we generate the matrix $J_{i j}$ from $A_{i j}$
of a network with desired properties,
by replacing the non-zero elements of the latter matrix with entries drawn
from a Gaussian distribution having mean zero and 
variance $\sigma^2$.
Note that, as the resultant 
matrices are not symmetric, the eigenvalues are expected to be found in 
the complex plane.
In this approach, correlations between the matrix elements are determined 
by the structure of the underlying graphs. 
In particular, third and higher order correlations are related to the
network assortativity, 
$$
a = \frac{\frac{1}{L} (\sum_{i=1}^L j_i k_i) 
- (\frac{1}{L} \sum_{i=1}^L \frac{1}{2} (j_i+k_i))^2}
        {\frac{1}{L} (\sum_{i=1}^L \frac{1}{2} (j_i^2+k_i^2) 
	- (\frac{1}{L} (\sum_{i=1}^L \frac{1}{2} (j_i+k_i)))^2},
$$
where $j_i, k_i$ are the degrees of nodes connected by the $i$-th link 
and $L \simeq N(N-1)C$ is the total number of
links in the network.

\begin{figure} [tbp]
\begin{center}
  \includegraphics [width=7cm]{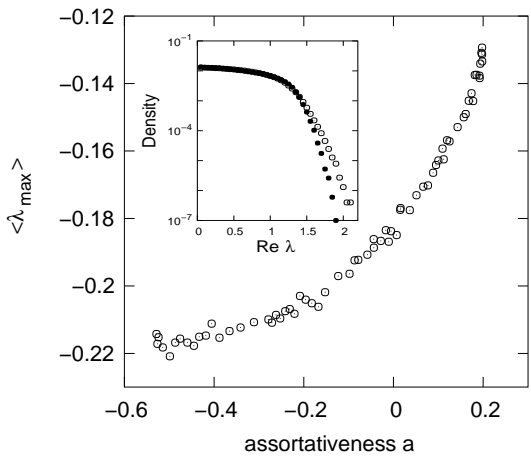}
 \caption{Dependence of the average of largest real part of the eigenvalues,
 $\langle \lambda_\text{max}\rangle$,
 on assortativity
 for Poisson degree distributed 
 networks of size $N=100$. 
 Data points represent averages over 
 $1000$ independent runs. The inset shows the tail of the real axis 
 projection of the spectrum for random matrices derived from 
 non-assortative ($a\approx 0$, filled circles) and strongly 
 assortative ($a=0.8$, open circles) networks. 
 The tail becomes more extended for strongly assortative networks. }
\label{F2}
\end{center}
\vspace{-1cm}
\end{figure}

First, we consider matrix ensembles constructed from graphs 
with Poisson degree distributions. 
We start with an ER graph (for which $a\approx 0$) 
and then introduce assortativity by applying the method proposed in 
Ref.~\cite{Brunet}. The algorithm 
essentially consists of performing a series of rewiring procedures, such that,
to obtain an assortative network, the
steps leading to higher (lower) assortativity are accepted with 
increased (decreased) probability.
Fig. \ref{F2} illustrates the dependence 
of $\langle \lambda_\text{max}\rangle$ on the assortativity.
These data show a clear trend towards 
larger $\langle \lambda_\text{max}\rangle$, i.e. less stability, 
as networks become more assortative. 
Note that, changing the assortativity by this rewiring procedure
also increases the
clustering coefficient, $c$. To correct for this, we use
the same scheme to change the clustering coefficient,
while keeping the degree distribution fixed. As this leaves the
assortativity unchanged, we can generate networks with
a specific value of $c$ while keeping $a$ constant.
In this case, there is no statistically significant change 
of  $\langle \lambda_\text{max}\rangle$ with $c$ (data not shown). 
Combining this result with Fig. \ref{F2} allows us to 
conclude that high assortativity leads to lower stability
for networks with Poisson degree distribution.

Next, we look at random matrices constructed from scale-free graph 
substrates. The networks are generated by adding at each time step a new node
with $m$ links that connect to existing nodes following the 
Barab\'{a}si-Albert (BA) preferential
scheme presented in 
Ref. \cite{Barabasi}. To avoid `dynamic correlations' resulting 
from the BA procedure, we subsequently randomise the networks using the 
link swapping technique discussed in Ref. \cite{Sneppen}. Corresponding
$J_{i j}$ matrices are constructed as before.
\begin{figure} [tbp]
\begin{center}
 \includegraphics [width=8cm]{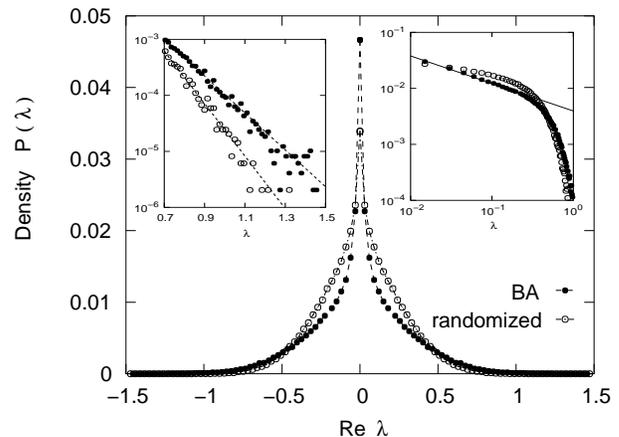}
 \caption{(Top) Real axis projection of the spectral densities 
 (shifted to the right by $\Delta \lambda=1$) for matrices constructed 
 from BA networks (filled circles) and randomised BA networks (open circles) 
 with $N=5000$, $m=10$, and $\sigma=0.2$. 
 The left inset illustrates the exponential tail of the spectrum, while
 the right inset shows that the spectrum follows a power 
 law $P(\lambda)\sim \lambda^{-1/2}$ for small $\lambda$ in the BA case, 
 but not after randomisation. The comparison illustrates 
 that the `dynamic correlations' in the BA network have an effect on the 
 spectrum. 
 The data represent averages over $100$ independent runs.}
\label{F5}
\end{center}
\vspace{-1cm}
\end{figure}
In Fig. \ref{F5}, a comparison of numerical data for the spectral 
densities of unrandomized and randomised BA networks demonstrates that the 
dynamical correlations in the former indeed have an impact on the 
spectral density. While both spectral densities have 
exponentially decaying tails [Fig.~\ref{F5} (left inset)], only the spectral 
density for the matrix ensemble constructed from unrandomized BA 
networks seems to obey a power law $P(\lambda)\sim \lambda^{-1/2}$ for 
small $\lambda$ [Fig.~\ref{F5} (right inset)]. 
Note that, this result obtained for weighted, directed graphs differs
significantly from the eigenvalue spectrum of adjacency matrices
of scale-free networks, which exhibit power-law tails and exponential decay
(or ``triangle-like'' shape) at the centre \cite{Goh,Farkas}.

%
We now observe the scaling of 
$\langle \lambda_\text{max}\rangle$ with network size $N$. 
The well known 
results for ER graphs suggest that $\lambda_\text{max}$ essentially is a 
function of the number of links in the network and their average link 
strength. Hence, to explore the dependence of 
$\langle \lambda_\text{max}\rangle$ on network size, 
we look at two cases. 

In the first case, as seems realistic for,
e.g., food webs \cite{Martinez}, we keep the connectivity $C$ of 
the networks constant as we increase the network size $N$. 
It follows that the 
number of links per node added, $m$, becomes a linear function of network size. 
\begin{figure} [tbp]
\begin{center}
\includegraphics [width=8cm]{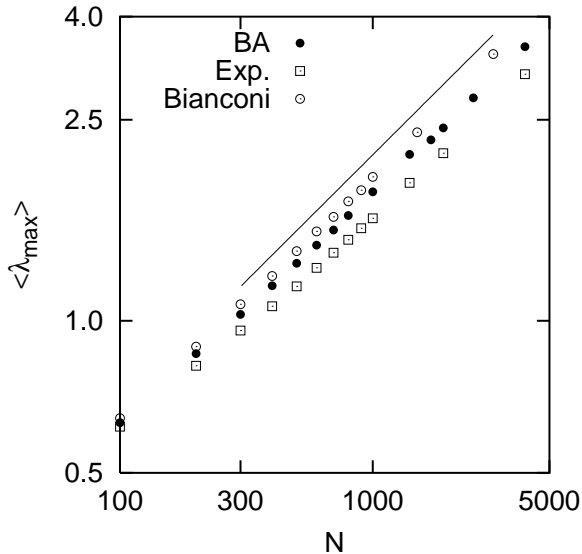}
\caption{Scaling of $\langle \lambda_\text{max}\rangle$ with
$N$ for exponential (open squares), BA (filled circles), and 
Bianconi (open circles) networks with constant connectivity $C=0.02$. 
The line indicates a power law with 
exponent $1/2$, implying that, in the asymptotic limit,
$\lambda_\text{max}\sim N^{1/2}$ for all three types of networks.}
\label{F6}
\end{center}
\vspace{-1cm}
\end{figure}
Fig. \ref{F6} shows the scaling of the average of largest real part of the
eigenvalues, $\langle \lambda_\text{max}\rangle$, with network size $N$. 
For three types of growing networks, viz., having exponential (i.e., no
preferential attachment) and scale-free degree distribution with 
exponents $\gamma=-3$ (the original BA preferential attachment algorithm) 
and $\gamma\approx -2.1$ 
(Bianconi networks, constructed after a modification of the BA procedure 
following 
Ref. \cite{Bianconi}), we find the same scaling behaviour as in ER graphs, 
$\lambda_\text{max} \sim N^{1/2}$, which
contrasts with the relation $\lambda_1\sim N^{1/4}$ 
reported for adjacency matrices of 
large undirected BA networks \cite{Goh}.
Variation among different network topologies are chiefly expressed 
in different finite size behaviour and proportionality constants.

In general, it seems that networks with degree 
distributions having a less extended tail are more stable. This appears to 
be supported by 
Ger$\hat{\text{s}}$gorins theorem \cite{Brualdi} which gives the
bound: $\lambda_\text{max} < R_\text{max}-1$, 
where $R_\text{max}=\max_i \sum_{j, j\not= i} |J_{i j}|$ is 
related essentially to the largest degree, $k_\text{max}$, of the network.
From
$ N\sum_{k_\text{max}}^{\infty} k^{-\gamma} \leq 1,$ 
where $k$ are node degrees (see e.g., Ref. \cite{Doro01}),
one obtains the following dependence of $k_\text{max}$ on $N$ for
the different networks: 
$k_\text{max} \sim N^{0.91}$ 
(for the Bianconi network), $k_\text{max} \sim N^{1/2}$ (for the BA 
network) and $k_\text{max} \sim \log (N)$ (for the exponential network). 
This ordering according to decreasing growth rates of $k_\text{max}$ with $N$
agrees very well with the 
ordering of the network types according to decreasing stability
in Fig. \ref{F6}.
\begin{figure} [tbp]
\begin{center}
 \includegraphics [width=7cm]{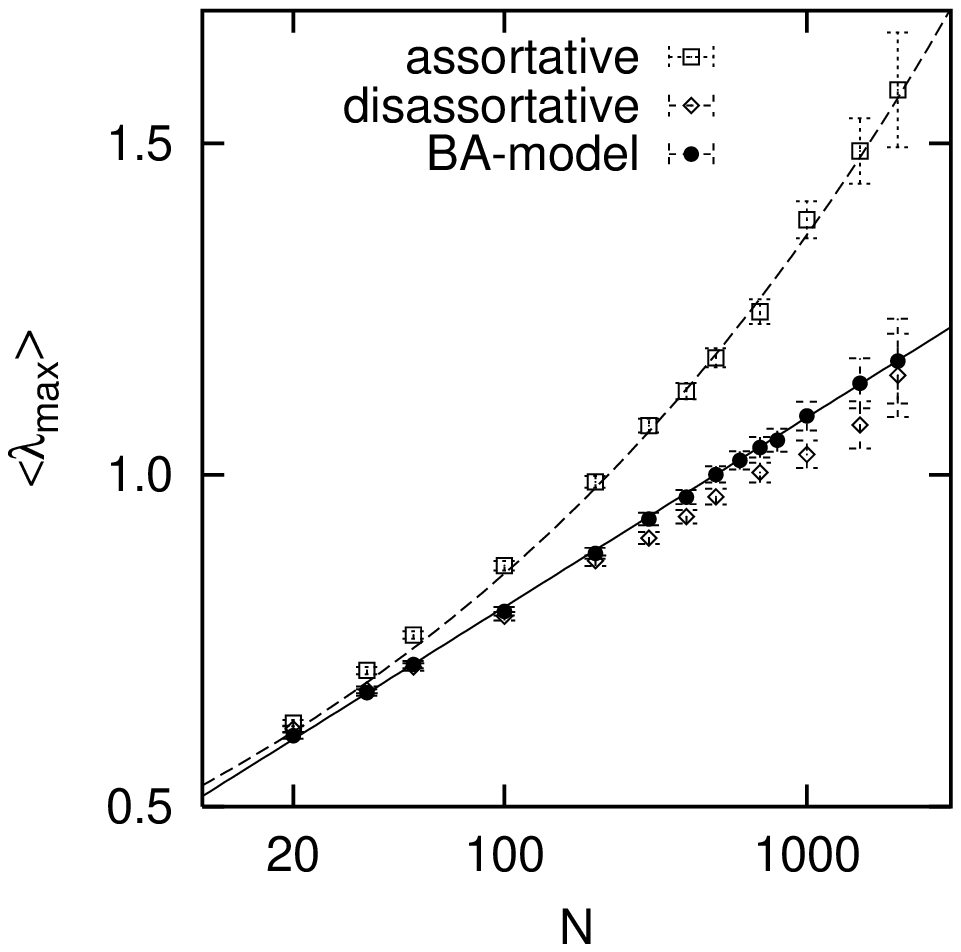}
  \includegraphics [width=7cm]{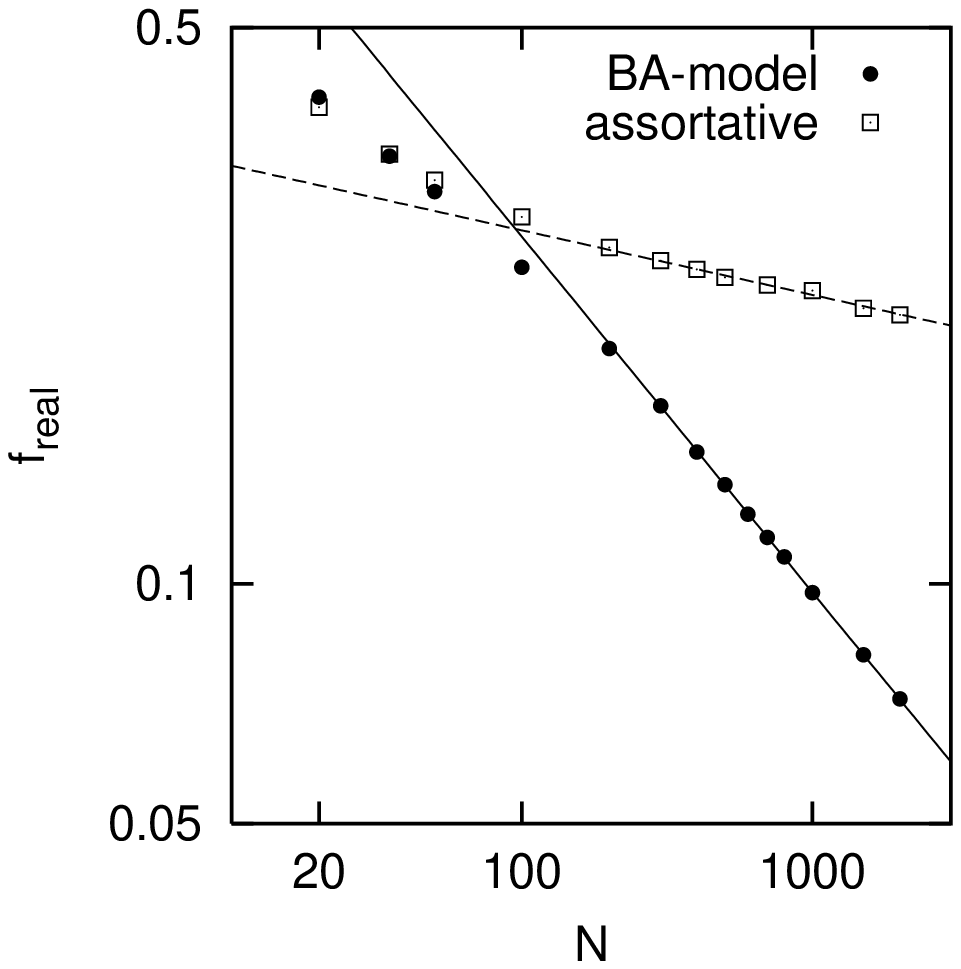}
\caption{(Top) Scaling of $\langle \lambda_\text{max}\rangle$ with 
network size $N$ when connectivity $C$ decreases with increasing $N$.
Data for networks constructed from the BA procedure (filled circles) 
are compared with networks optimised for assortativity
(open boxes, $a\approx 0.24$, $c\approx 0.45$) and for 
disassortativity (open diamonds, $a\approx -0.2$, $c=0$). For the 
assortative networks,
$ \lambda_\text{max}\sim N^{0.2}$ (dashed line), while 
for disassortative and 
uncorrelated BA networks, 
$\lambda_\text{max}\sim \log(N^{0.12})$ (solid line) for 
large $N$. (Bottom) Scaling of the excess density $f_\text{real}$ 
of real eigenvalues with $N$: 
$f_\text{real}\propto N^{- 0.1}$ (dashed line) for the assortative networks, 
and $f_\text{real} \propto N^{- 0.45}$ (solid line) for uncorrelated BA and 
disassortative networks. Data points represent averages over 
$1000$ runs ($N<800$) and $200$ runs ($N\geq 800$).}
\label{F7}
\end{center}
\vspace{-1cm}
\end{figure}

As already mentioned, in the preceding case we considered the network
connectivity $C$ to be constant as $N$ increases.
However, most real networks are sparse, i.e. $C\sim N^{-1}$ 
(see, e.g., Ref.~\cite{BarabRev}). Therefore, in the second case, we measure  
$\langle \lambda_\text{max}\rangle$ in networks with increasing size, 
while keeping $m$ constant (resulting in decreasing connectivity $C$
with increasing $N$). 
We now compare the behaviour of $\lambda_\text{max}$ for assortative and 
disassortative networks, constructed using the same method as before. 
For the results reported in this paper, we choose $m=2$, for which we verify
numerically that, after randomisation, the resulting networks are still
connected.
Optimising for a high assortativity, 
however, tends to fracture this component, e.g., we find that networks 
with $a\approx 0.25$ typically have a giant component that 
comprises only $80\%$ of the links in the network. 
Note that, this fraction of links comprising the largest connected cluster, is 
independent of the number of nodes $N$.

Fig.~\ref{F7}~(top) shows the scaling of 
$\langle \lambda_\text{max}\rangle$ with $N$, for constant $m$. 
As in the case of the 
ER graph-based matrices, we again find that high assortativity leads to 
lower stability. For the assortative scale-free networks with $a\approx 0.25$ 
we find that $\lambda_\text{max} \sim N^{0.2}$, whereas for the uncorrelated
and disassortative networks,
$\lambda_\text{max}\sim \log(N^{0.12})$ for large $N$. 
This difference in the scaling behaviour is suggestive of a huge difference
in stability for large 
network sizes. 

The reason for the different scaling of $\lambda_\text{max}$ 
in assortative and disassortative networks seems to be found in
different scaling behaviours of the excess densities $f_\text{real}$ of 
real eigenvalues [Fig.~\ref{F7} (bottom)].
For random matrices based on a ER graph topology, it is known
that $f_\text{real}\propto N^{-1/2}$ for large $N$ \cite{Sommers}.
For uncorrelated BA networks we find that
$f_\text{real} \propto N^{-\delta}$, $\delta=0.45 \pm 0.05$, 
which is very close to 
the value for ER graphs. However, for assortative networks, we observe
$\delta=0.1\pm 0.05$, which results in a much more gradual decline than 
in disassortative networks. Hence, for the same 
network size $N$, more eigenvalues lie on the real axis in 
assortative networks and thus the probability of having higher maximum real 
parts is increased. 
We also note that, as in the case of matrices based on graphs
with Poisson degree distribution
[Fig.~\ref{F2} (inset)], the tails at the spectrum edge show 
a more gradual decay for assortative networks,
compared to uncorrelated and disassortative graphs (not shown).

In this paper, we have presented an analysis of randomly weighted matrices 
whose corresponding networks are 
constructed by starting from well known graph models such as
ER and scale-free networks, and then using a link swapping algorithm
to generate 
degree-correlated network ensembles with a desired assortativity.
As the most important result of the paper, we find that
assortative mixing by degree tends to severely destabilise networks. 
While the largest 
real part of the eigenvalues scales as 
a power of the network size in highly assortative networks, in the 
disassortative case we only observe a logarithmic dependence. 

The question has been raised why almost all biological networks are 
disassortatively mixed \cite{Agenda}. 
It has been suggested that disassortativity has its origin in the restriction
that no two nodes have more than one link between them. However, a recent study
\cite{Park03} shows that while this mechanism does indeed give rise to the
kind of correlations actually observed in certain networks (e.g., the Internet),
{\em only a part} of all the measured correlations can be accounted in this 
way.
Other studies have pointed out
that assortative mixing has a big impact on the percolation 
behaviour of networks \cite{Brunet}. This implies that disassortative 
networks might have been favoured because perturbations have a reduced 
chance to propagate. 
Our results add another viewpoint to this issue. The results reported here
imply that disassortative networks are more resistant to the effect of 
dynamical fluctuations
than assortative networks \cite{Sorrentino}. 
One may therefore speculate that, an evolutionary 
drive towards systems 
with reduced intensity of fluctuation (homeostasis) might be one of the 
reasons for the prevalence of disassortative networks in the 
biological world today.


\begin{thebibliography}{99}
\bibitem {McCann} {K. S. McCann, Nature (London) {\bf 405}, 228 (2000).}
\bibitem {May1} {R. M. May, Phil. Trans. Roy. Soc. B {\bf 354}, 1951 (1999).}
\bibitem {May2} {R. M. May, {\it Stability and complexity in model 
ecosystems}, (Princeton University Press, Princeton, NJ, 1973).}
\bibitem {Sommers} {H. J. Sommers, A. Crisanti, H. Sompolinsky, and Y. Stein, 
Phys. Rev. Lett. {\bf 60}, 1895 (1988).}
\bibitem{Mehta}{M. L. Mehta, {\it Random matrices}, 2nd ed. (Academic, 
San Diego, 1991). }
\bibitem {Jirsa} {V. K. Jirsa and M. Ding, Phys. Rev. Lett. {\bf 93},
070602 (2004).} 
\bibitem {Sin1} {S. Sinha and S. Sinha, Phys. Rev. E {\bf 71}, 020902(R)
(2005).}
\bibitem {Sin2} {S. Sinha, Physica A {\bf 346}, 147 (2005).}
\bibitem {Erdos} {P. Erd\"{o}s and A. R\'{e}nyi, Publ. Math. Inst. Hung. 
Acad. Sci., Ser. A {\bf 5}, 17 (1960).}
\bibitem {BarabRev} {R. Albert and A.-L. Barab\'{a}si, 
Rev. Mod. Phys. {\bf 74}, 47 (2002).}
\bibitem {NewmanRev} {M. E. J. Newman, SIAM Review {\bf 45}, 167 (2003).}
\bibitem {Barabasi} {A.-L. Barab\'{a}si and R. Albert, Science {\bf 286}, 
509 (1999).}
\bibitem {Goh} {K.I. Goh, B. Kahng, and D. Kim, Phys. Rev. E {\bf 64}, 
051903 (2001).}
\bibitem {Farkas} {I.J. Farkas, A.-L. Barab\'{a}si and T. Vicsek, 
Phys. Rev. E {\bf 64}, 026704 (2001).}
\bibitem {Dorogovtsev} {S. N. Dorogovtsev, A. V. Goltsev, J. F. F. Mendes,
and A. N. Samukhin, Phys. Rev. E {\bf 68}, 046109 (2003).}
\bibitem {Newman02} {M. E. J. Newman, Phys. Rev. Lett. {\bf 89}, 
208701 (2002).}
\bibitem {Newman03} {M. E. J. Newman, Phys. Rev. E {\bf 67}, 026126 (2003).} 
\bibitem {Agenda} {Eur. Phys. J. B {\bf 38}, 143 (2004).}
\bibitem {Brunet} {R. Xulvi-Brunet and I. M. Sokolov, 
Phys. Rev. E {\bf 70}, 066102 (2004).}
\bibitem {Sneppen} {S. Maslov and K. Sneppen, 
Science {\bf 296}, 910 (2002).}
\bibitem {Martinez} {N. D. Martinez, Am. Nat. {\bf 139}, 1208 (1992).}
\bibitem {Bianconi} {G. Bianconi and A.-L. Barab\'{a}si, 
Eur. Phys. Lett. {\bf 54}, 436 (2001).}
\bibitem{Brualdi}{R. A. Brualdi and H. J. Ryser, {\it Combinatorial Matrix 
Theory} (Cambridge University Press, Cambridge, 1991). }
\bibitem{Doro01} {S. N. Dorogovtsev, J. F. F. Mendes and A. N. Samukhin,
Phys. Rev. E {\bf 63}, 062101 (2001).}
\bibitem{Park03} {J. Park and M. E. J. Newman, Phys. Rev. E {\bf 68}, 026112
(2003).}
\bibitem{Sorrentino} {Our conclusions are supported by recent results
that disassortative mixing enhances synchronizability of directed weighted 
networks. See, e.g., M. di Bernardo, F. Garofalo and F. Sorrentino,
cond-mat/0504335.}
\end{thebibliography}
\end{document}